\documentclass[a4paper,11pt]{article}
\usepackage{pos}

\usepackage{braket}
\usepackage{xargs}
\usepackage{mathtools}
\usepackage[english]{babel}
\usepackage[utf8]{inputenc}
\usepackage{amsmath}
\usepackage{pifont}
\usepackage{MnSymbol}
\usepackage{graphicx}
\usepackage{tikz}
\usetikzlibrary{shapes,arrows,positioning}
\tikzset{decision/.style={diamond, draw, fill=blue!20, text width=4.5em, text badly centered, inner sep=0pt}}
\tikzset{block/.style={rectangle, draw, fill=blue!20, text width=10em, text centered, rounded corners, minimum width=3.5cm}}
\tikzset{block1/.style={rectangle, draw, fill=blue!20, text width=18.5em, text centered, rounded corners, minimum width=3.5cm}}
\tikzset{line/.style={draw, -latex, thick}}
\usepackage{color,hyperref}
\usepackage{multirow,array}
\usepackage{physics}
\usepackage{cleveref}
\newcommand{\be}{\begin{equation}}
\newcommand{\ee}{\end{equation}}
\newcommand{\nn}{\nonumber}
\newcommand{\ud}{\mathrm{d}}

\newcommand{\pkuphy}{School of Physics, Peking University, Beijing 100871, China}

\title{Lattice QCD calculation of the two-photon exchange contribution to the muonic-hydrogen Lamb shift}
\ShortTitle{Two-photon exchange contribution to the $\mu$H Lamb shift}

\author*[a]{Yang Fu}
\affiliation[a]{\pkuphy}

\emailAdd{fy\_deg@pku.edu.cn}

\abstract{
	We develop a method for lattice QCD calculation of the two-photon exchange (TPE) contribution to the muonic-hydrogen Lamb shift. To demonstrate the feasibility of this method, we also present an exploratory study with a gauge ensemble at $m_\pi = 142$ MeV. By adopting the infinite-volume reconstruction (IVR) method along with an optimized subtraction scheme, we obtain a preliminary result of the TPE contribution which agrees well with previous calculation using other methods and one magnitude smaller compare to the large $\sim300~\mu$eV discrepancy for the proton radius puzzle.
}

\FullConference{
	The 38th International Symposium on Lattice Field Theory, LATTICE2021
	26th-30th July, 2021
	Zoom/Gather@Massachusetts Institute of Technology
}

\begin{document}
\maketitle

\section{Introduction}
\begin{figure}[htbp]
		\begin{center}
			\includegraphics[width=50mm]{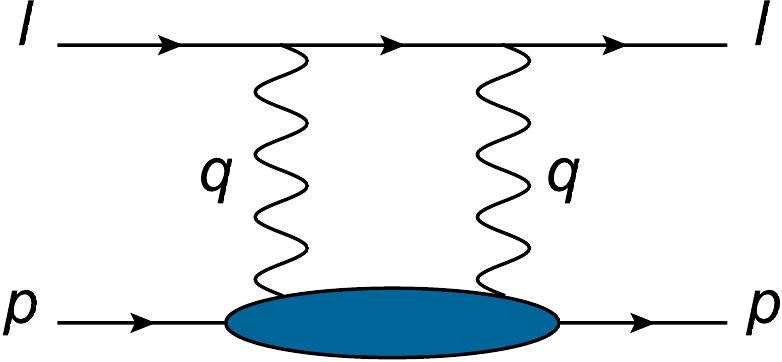}
			\caption{The diagram for two-photon exchange correction.}
			\label{fig:TPE}
		\end{center}
	\end{figure}

	The measurement of muonic-hydrogen spectroscopy \cite{Pohl:2010zza,Antognini:2013txn} not only provides the most precise determination of the proton charge radius, but also raises the unexpected proton radius puzzle. This puzzle triggers a great deal of efforts to improve the theoretical corrections to both spectroscopy and scattering. Among them, the two-photon exchange (TPE) correction, see Fig.~\ref{fig:TPE}, is of special interest. It involves a wealth of information about the proton structure and introduces the largest theoretical uncertainty to both the Lamb shift and hyperfine splitting in muonic-hydrogen \cite{Antognini:2013rsa}. The TPE correction also plays an important role in the electron-proton scattering since it could be responsible for the drastic difference in the ratio of the proton electric to magnetic form factors obtained using the Rosenbluth separation \cite{Rosenbluth:1950yq} and the polarization transfer methods \cite{JeffersonLabHallA:1999epl}. 
	
	Several approaches have been applied in previous work, including dispersion relations (DR) \cite{Pachucki:1999zza,Martynenko:2005rc,Carlson:2011zd,Gorchtein:2013yga,Tomalak:2018uhr}, baryon $\chi$PT (B$\chi$PT) \cite{Alarcon:2013cba,Alarcon:2020wjg}, heavy baryon $\chi$PT (HB$\chi$PT) \cite{Nevado:2007dd,Birse:2012eb,Peset:2015zga}, non-relativistic QED (NRQED) \cite{Hill:2011wy} and operator product expansion (OPE) \cite{Hill:2016bjv}. For these methods, the TPE correction is usually divided into Born and non-Born pieces, where as in the DR approach, the Born part can be well-constrained by the experiment data, but the non-Born part contains a component, commonly referred as "subtraction function", which is poorly constrained and relies on model, thus leading to a large systematic uncertainty. Other theoretical approaches are usually devoted to improve the determination of the non-Born contribution, in particular the contribution of subtraction function. It has also been recently proposed that the subtraction function can be further constrained by the dilepton electroproduction \cite{Pauk:2020gjv}.
	
	The total $2S-2P$ Lamb shift in muonic hydrogen is given by \cite{Antognini:2013rsa} (units in meV and fm)
	\be \label{eq:muH_LS_theory}
	\Delta E_\text{LS}^{\text{theory}} = 206.0336(15) - 5.2275(10) \langle r_p^2 \rangle + \Delta E_\text{TPE},
	\ee
	with their choice of TPE contribution $\Delta E_\text{TPE} = 0.0332(20)$~meV \cite{Carlson:2011zd,Birse:2012eb} and experiment result $\Delta E_\text{LS}^{\text{exp}} = 202.3706(23)$ meV \cite{Antognini:2013txn}. One find that the uncertainty of the TPE is at the same level as the present experimental precision, thus any further improvement on proton charge radius extraction from $\mu$H Lamb shift will unavoidably require an improved TPE determination, of which the precise lattice QCD calculation is undoubtedly important.
	
	Several lattice QCD approaches have been recently proposed, including using the Feynman-Hellmann theorem to calculate the structure function \cite{Can:2020sxc} and using an unconventional choice of the subtraction point to calculate the non-Born contribution of the TPE \cite{Hagelstein:2020awq}. In this work, we develop a method to directly calculate the TPE correction to the $\mu$H Lamb shift on the lattice. We also perform an exploratory study with a gauge ensemble at the physical pion mass. The preliminary result is consistent with previous data-driven analysis.
	
\section{Two-photon exchange contribution}
	We start with the spin-averaged forward doubly-virtual Compton scattering (VVCS) tensor in Euclidean space. With $H_{\mu \nu}(x) = \langle p \vert \mathcal{T}[j_\mu(x) j_\nu(0)] \vert p \rangle$, we have
	\begin{align} \label{eq:Compton_tensor}
		T_{\mu\nu}(P,Q) &= \frac{1}{8\pi M} \int \ud^4 x e^{iQ \cdot x} H_{\mu \nu}(x)\nn \\[1ex]
		& = \left( -\delta_{\mu \nu} + \frac{Q_\mu Q_\nu}{Q^2} \right)  T_1(\nu, Q^2) - \left( P_{\mu} - \frac{P \cdot Q}{Q^2} Q_{\mu}\right)\left( P_{\nu} - \frac{P \cdot Q}{Q^2} Q_{\nu}\right) \frac{T_2(\nu, Q^2)}{M^2} ,
	\end{align}
	where $\nu = P\cdot Q/M$ with $P$ and $Q$ the Euclidean proton and photon four-momenta, and $M$ is the proton mass. 
	
	For the TPE contribution, the Euclidean momenta are chosen as $P = (iM,\vec{0})$ and $Q = (Q_0,\vec{Q})$. The relative energy shift to the the $nS$-state is then given by \cite{Pachucki:1999zza}
	\begin{align} \label{eq:TPE}
		\Delta E &=  \frac{8 m \alpha^2}{ \pi } \abs{\phi_n(0)}^2 \int \ud^4Q \frac{ (Q^2+2Q_0^2) T_1(iQ_0,Q^2) - (Q^2 -Q_0^2) T_2(iQ_0,Q^2) }
		{ Q^4 (Q^4 + 4m^2 Q_0^2) }	.
	\end{align}
	where $m$ is the lepton mass, $\abs{\phi_n^2(0)} = {m_r^3 \alpha^3}/{(\pi n^3)}$ is the square of the $nS$-state wave function
	at the origin with $m_r = mM/(M+m)$ the reduced mass. Note that the $nP$-state wave function vanishes at the origin, hence it won't receive any corrections from TPE at this order. Eq.(\ref{eq:TPE}) essentially contains an infrared singularity which is due to the terms already accounted for at the lower order \cite{Carlson:2011zd}: one is the point-like proton contribution, which means the proton-photon vertex function $\Gamma_\mu = \gamma_\mu$, hence
	\be \label{eq:sub1}
	T_{1}^{(\text{pt})} = \frac{M}{\pi} \frac{\nu^2}{Q^4 - 4M^2 \nu^2},\quad T_{2}^{(\text{pt})} =  \frac{M}{\pi} \frac{Q^2}{Q^4 - 4M^2 \nu^2} ,
	\ee
	another one is the charge radius term from third Zemach moment contribution
	\be \label{eq:sub2}
	\Delta E^{(\text{Z})} = - \alpha^2 \abs{\phi_n(0)}^2 \int \frac{\ud Q^2}{Q^2} \frac{16mM}{(M+m)Q} G_E^\prime(0) ,
	\ee
	with $G_E(Q^2)$ the proton electric form factor and its derivative can be related to proton charge radius via $\langle r_p^2 \rangle = -6 G_E^\prime(0)$. These terms should be subtracted in order to both keep the TPE contribution IR finite and avoid double-counting.
	
\section{Lattice QCD methodology}
	On the lattice, we prefer to rewrite Eq.(\ref{eq:TPE}) in terms of $T_{00}$ and $\sum_i T_{ii}$
	\be \label{eq:TPE_lat}
	\Delta E = -16 m \alpha^2 \abs{\phi_n(0)}^2 \int \frac{\ud Q^2}{Q^2} \int_{-\frac{\pi}{2}}^{\frac{\pi}{2}} \ud \theta \left( g_1 T_{00} + g_2 \sum_i T_{ii} \right) ,
	\ee
	with
	\be
	g_1 = \frac{1 - \sin^4\theta}{Q^2 + 4m^2 \sin^2 \theta}, \quad g_2 = \frac{\sin^2\theta (1 - \sin^2\theta)}{Q^2 + 4m^2 \sin^2 \theta} ,
	\ee
	and the angle $\theta$ defined as
	\be \label{eq:angle}
	Q_0 = Q \sin\theta, \quad \abs{\vec{Q}} = Q \cos\theta .
	\ee
	The point-like proton contribution from Eq.(\ref{eq:sub1}) can also be represented in terms of $T_{00}$ and $\sum_i T_{ii}$
	\be \label{eq:sub1_lat}
	T_{00}^{(\text{pt})} = \frac{M}{\pi} \frac{Q^2 - Q_0^2}{Q^4 + 4M^2 Q_0^2},\quad \sum_i T_{ii}^{(\text{pt})} = \frac{M}{\pi} \frac{3Q_0^2}{Q^4 + 4M^2 Q_0^2} .
	\ee
	Combining Eq.(\ref{eq:Compton_tensor}) and (\ref{eq:TPE_lat}), we obtain
	\be \label{eq:master_formula_naive}
	\Delta E = \frac{2 m \alpha^{2}}{\pi M} \abs{\phi_n(0)}^2 \sum_{i=1,2} \int \ud^{4} x ~ \omega_{i}(\vec{x}, t) H_{i}(\vec{x}, t) ,
	\ee
	here the hadronic functions are defined as $H_{1}(\vec{x}, t) = H_{00} (\vec{x}, t)$ and $H_{2}(\vec{x}, t) = \sum_i H_{ii} (\vec{x},t)$, with weight functions given by
	\be
	\omega_{i}(\vec{x}, t) = - \int \frac{\ud Q^2}{Q^2} \int_{-\frac{\pi}{2}}^{\frac{\pi}{2}} \ud \theta f(Q;x) g_i , \quad f(Q;x) = \cos(Q_0 t) j_0(\abs{\vec{Q}} \abs{\vec{x}}),
	\ee
	where an average over the spatial directions is taken and $j_n(x)$ are the spherical Bessel functions. One immediately find these weight functions are IR divergent since we have not performed the subtraction.
	
	The IR divergence occurs only in the elastic contribution, then both of the two terms given by Eq.(\ref{eq:sub2}) and (\ref{eq:sub1_lat}) can be reproduced by the ground-state contribution on the lattice. Choosing a sufficiently large time $t_s$ for the ground-state saturation, we obtain that
	\begin{align} \label{eq:GS}
		\tilde{H}_{i}(\vec{Q},t_s) =& \int \ud^3 \vec{x} ~ j_0(\abs{\vec{Q}} \abs{\vec{x}}) H_{i}(\vec{x},t_s) \nn \\[1ex]
		=& \frac{M}{E_Q} e^{-(E_Q-M)t_s} \times \left\{ 
		\begin{aligned}
			&(E_Q+M) G_E^2(Q^2_{\text{on}}), && i=1,\\
			&- (E_Q-M) [G_E^2 + 2G_M^2](Q^2_{\text{on}}), && i=2,
		\end{aligned}\right.
	\end{align}
	here $G_E$, $G_M$ are the proton electric and magnetic form factors, with $E_Q = \sqrt{M^2+\vec{Q}^2}$ the proton energy and $Q^2_{\text{on}} = 2M(E_Q-M)$. The low-momentum expansion of Eq.(\ref{eq:GS}) gives that
	\be \label{eq:func_G_E}
		G_E^2(0) = \int \ud^3 \vec{x} L_0(\vec{x},t_s) H_{1}(\vec{x},t_s) , \quad
		\langle r_p^2 \rangle = \int \ud^3 \vec{x} L_r(\vec{x},t_s) H_{1}(\vec{x},t_s) ,
	\ee
	with
	\be \label{eq:sub_weight_func}
	L_0(\vec{x},t_s) = \frac{1}{2M},\quad L_r(\vec{x},t_s) = \frac{1}{4M} \left(x^2 - \frac{3+6M t_s}{2M^2} \right).
	\ee
	The similar idea has been applied to the pion electromagnetic transition to extract the charge radius~\cite{Feng:2019geu}.
	The terms need to be subtracted can then be given by
	\be
	\Delta E^{(\text{sub})} = \frac{2 m \alpha^{2}}{\pi M} \abs{\phi_n(0)}^2 \int \ud^{3} \vec{x} ~ L^{(\text{sub})}(\vec{x},t_s) H_{1}(\vec{x}, t_s), \nn \\[1ex]
	\ee
	with three different weight functions $L^{(\text{sub})}(\vec{x},t_s)$
	\begin{align}
		L_{00}^{(\text{pt})}(\vec{x},t_s) &= - \int \frac{\ud Q^2}{Q^2} \int_{-\frac{\pi}{2}}^{\frac{\pi}{2}} \ud \theta S_1(Q) g_1 , \nn \\[1ex]
		L_{ii}^{(\text{pt})}(\vec{x},t_s) &= - \int \frac{\ud Q^2}{Q^2} \int_{-\frac{\pi}{2}}^{\frac{\pi}{2}} \ud \theta S_2(Q) g_2, \nn \\[1ex]
		L^{(\text{Z})}(\vec{x},t_s) &= \int \frac{\ud Q^2}{Q^2} \frac{4\pi M^2}{3(M+m)Q}L_r(\vec{x},t_s).
	\end{align}
	here
	\be
		S_1(Q) = 8\pi M T_{00}^{(\text{pt})} L_0(\vec{x},t_s) = \frac{4M (Q^2-Q_0^2)}{Q^4 + 4M^2 Q_0^2}, \quad
		S_2(Q) = 8\pi M \sum_i T_{ii}^{(\text{pt})} L_0(\vec{x},t_s) = \frac{12M Q_0^2}{Q^4 + 4M^2 Q_0^2}.
	\ee
	The subtraction of the hadronic matrix elements has therefore been transferred to the subtraction of weight functions, which facilitates the lattice QCD calculation.
	
	We further adopt the infinite-volume reconstruction (IVR) method~\cite{Feng:2018qpx}, which is developed to
remove all the power-law finite-volume effects in the QED self-energy~\cite{Feng:2021zek}. 
This method has been successfully applied to the lattice study of double beta decays~\cite{Tuo:2019bue}, rare decays~\cite{Christ:2020hwe} and the leptonic decays~\cite{Christ:2020jlp,Tuo:2021ewr}. In the work, the IVR method plays a crucial role in the removal of the infrared divergence. In practise,
the time integral in Eq.(\ref{eq:Compton_tensor}) is split into the range of $\abs{t}<t_s$ and $\abs{t}>t_s$
	\be \label{eq:Compton_tensor_IVR}
		T_{\mu\nu} = \frac{1}{8\pi M} \bigg[ \int_{\abs{t}<t_s} \ud^4 x ~ f(Q;x) H_{\mu \nu}(\vec{x}, t) + \int \ud^{3} \vec{x} ~ S(Q;\vec{x}, t_s) H_{\mu \nu}(\vec{x}, t_s) \bigg],
	\ee
	with
	\begin{align}
		S(Q;\vec{x}, t_s) = &~2 \int \frac{\ud^3 \vec{Q}}{(2\pi)^3} e^{i \vec{Q} \cdot \vec{x}} \int_{t_s}^{\infty} \ud t ~ e^{-(E_Q-M)(t-t_s)} \int \ud^3 \vec{x}^\prime e^{- i \vec{Q} \cdot \vec{x}^\prime} f(Q;x) \nn \\[1ex]
		=&~ \frac{4M Q^2}{Q^4 + 4M^2 Q_0^2} A(Q;t_s) j_0(\abs{\vec{Q}} \abs{\vec{x}}) ,
	\end{align}
	here
	\be
	A(Q;t_s) = A_c (Q) \cos(Q_0 t_s) - A_s(Q) \sin(Q_0 t_s) ,
	\ee
	and the auxiliary functions are
	\be
		{A}_c(Q) = \sqrt{\frac{1}{4}+\tau_p \cos^2\theta} +\frac{1}{2} - \sin^2\theta , \quad {A}_s(Q) = \frac{\sin\theta}{\sqrt{\tau_p}}\left( \sqrt{\frac{1}{4}+\tau_p \cos^2\theta} +\frac{1}{2} + \tau_p \right).
	\ee
	with $\tau_p = Q^2/(4M^2)$ and the angle $\theta$ defined in Eq.(\ref{eq:angle}). Correspondingly, the time integral in $\Delta E$ can be also split into
	\begin{align} \label{eq:master_formula}
		\Delta E_{\abs{t}<t_s} &= \frac{2 m \alpha^{2}}{\pi M} \abs{\phi_n(0)}^2 \sum_{i=1,2} \int_{\abs{t}<t_s} \ud^{4} x ~ \omega_{i}(\vec{x}, t) H_{i}(\vec{x}, t), \nn \\[1ex]
		\Delta E_{\abs{t}>t_s} &= \frac{2 m \alpha^{2}}{\pi M} \abs{\phi_n(0)}^2 \sum_{i=1,2} \int \ud^{3} \vec{x} ~ L_{i}(\vec{x},t_s) H_{i}(\vec{x}, t_s),
	\end{align}
	here the weight functions $L_i(\vec{x},t_s)$ are given by
	\be
	L_i(\vec{x},t_s) = - \int \frac{\ud Q^2}{Q^2} \int_{-\frac{\pi}{2}}^{\frac{\pi}{2}} \ud \theta S(Q;\vec{x},t_s) g_i .
	\ee
	Currently both two types of weight functions $\omega_i(\vec{x},t)$ and $L_i(\vec{x},t_s)$ are IR divergent, but the subtraction is only performed to the latter. Considering that the $\Delta E$ should be finite, these weight functions will still be IR divergent even after performing the subtraction. The hadronic functions from $\abs{t}<t_s$ and $\abs{t}>t_s$, however, can be constrained by the low-energy expansion (LEX) of the VVCS tensor, thus are not completely unrelated. The LEX of $T_{\mu \nu}$ gives \cite{Gasser:2015dwa}
	\be 
	T_{\mu \nu}(Q) = T^{\text{Born}}_{\mu \nu}(Q) + \mathcal{O}(Q^2),
	\ee
	where the Born terms are standard \cite{Carlson:2011zd,Birse:2012eb,Gasser:2015dwa}. Comparing this LEX with Eq.(\ref{eq:GS}) and (\ref{eq:Compton_tensor_IVR}) in the low-$Q$ limit, we find that
	\be
	\frac{1}{8 \pi M} \int_{\abs{t}<t_s} \ud^4 x [H_i (\vec{x},t) - H_i (\vec{x},t_s)] = \left\{
	\begin{aligned}
		&0, &&i = 1,\\
		&\frac{3}{4 \pi M}, &&i = 2.
	\end{aligned}\right.
	\ee
	This relation ensures that the IR divergence of $\omega_i(\vec{x},t)$ and $L_i(\vec{x},t_s)$ can be exactly canceled out. We modify these weight functions to
	\be \label{eq:omegai}
	\omega_{i}(\vec{x}, t) = - \int \frac{\ud Q^2}{Q^2} \int_{-\frac{\pi}{2}}^{\frac{\pi}{2}} \ud \theta [f(Q;x) - 1] g_i ,
	\ee
	and
	\be
	L_i(\vec{x}, t_s) = - \int \frac{\ud Q^2}{Q^2} \int_{-\frac{\pi}{2}}^{\frac{\pi}{2}} \ud \theta [S(Q;\vec{x},t_s) + 2t_s] g_i ,
	\ee
	which does not change the contribution from temporal component $T_{00}$, but does remove a term of $3/(4\pi M)$ from the spatial component $\sum_i T_{ii}$ hence needs to be added back. Such term can be directly combined with the subtraction of $\sum_i T_{ii}^{(\text{pt})}$ defined in Eq.(\ref{eq:sub1_lat}) and leaves a term of ${\frac{3}{4\pi M} \left(\frac{Q^4}{Q^4-4M^2 \nu^2}\right)}$ which contributes a finite $-0.60~\mu$eV to the total TPE energy shift. Finally, after performing the subtraction of remaining contribution from $L_{00}^{(\text{pt})}(\vec{x},t_s)$ and $L^{(\text{Z})}(\vec{x},t_s)$, we obtain that
	\be \label{eq:Li}
		L_i(\vec{x}, t_s) = - \int \frac{\ud Q^2}{Q^2} \bigg\{\int_{-\frac{\pi}{2}}^{\frac{\pi}{2}} \ud \theta \Big([S- S_1](Q;\vec{x},t_s) + 2t_s \Big)g_i + \frac{4\pi M^2}{3(M+m)Q}L_r(\vec{x},t_s) \delta_{i,1} \bigg\} ,
	\ee
	with $S_1(Q;\vec{x},t_s) = S_1(Q)$. Here a trivial subtraction is also applied to $L_2(\vec{x}, t_s)$ to make it IR finite, since Eq.(\ref{eq:GS}) shows that $\int \ud^3 \vec{x} H_{2} (\vec{x},t_s) = 0$.
	
\section{Optimized subtraction scheme}
	Eq.(\ref{eq:master_formula}) along with the weight functions given by Eq.(\ref{eq:omegai}) and (\ref{eq:Li}) provides a direct way to calculate the TPE contribution using hadronic functions $H_i(\vec{x}, t)$ as input, but it suffers from both the finite-volume effects and the signal-to-noise problem in a realistic lattice QCD calculation, due to the fact that $L_1(\vec{x}, t_s)$ increases rapidly as the spatial distance $\abs{\vec{x}}$ increases. However, not all the contributions from $L_1(\vec{x}, t_s)$ need to be determined on the lattice. Inspired by Eq.(\ref{eq:muH_LS_theory}), We can divide the $\Delta E$ into
	\be
	\Delta E = - 0.60~\mu\text{eV} + c_0 + c_r \langle r_p^2 \rangle + \Delta E^{(\text{lat})},
	\ee
	by splitting the weight function $L_1(\vec{x},t_s)$
	\be \label{eq:L1_sub}
	L_1(\vec{x}, t_s) = \bar{c}_0 L_0(\vec{x}, t_s) + \bar{c}_r L_r(\vec{x}, t_s) + L_1^{(r)}(\vec{x}, t_s),
	\ee
	with $\bar{c} = c / \left(\frac{2 m \alpha^{2}}{\pi M} \abs{\phi_n(0)}^2 \right)$. As a result, only the $\Delta E^{(\text{lat})}$ need to be calculated on the lattice using the reduced weight function $L_1^{(r)}(\vec{x}, t_s)$ along with other three weight functions. The choice of two coefficients $c_0$ and $c_r$ can be viewed as a subtraction scheme. We choose them by minimizing the following integral
	\be
	\int_{R_\text{min}}^{R_\text{max}} \ud x ~ 4 \pi x^2 [L_1(x,t_s) - \bar{c}_0 L_0(x,t_s) - \bar{c}_r L_r(x,t_s)]^2.
	\ee
	with sufficiently large $t_s$ for ground state saturation, as well as the range $R_\text{min}$ to $R_\text{max}$ dominants the contribution of the integral. In this way, the long-distance contribution can be almost completely represented by the charge conservation and charge radius terms, which eliminates the need for lattice data as input.
	
	\begin{figure}[htbp]
		\begin{center}
			\includegraphics[width=100mm]{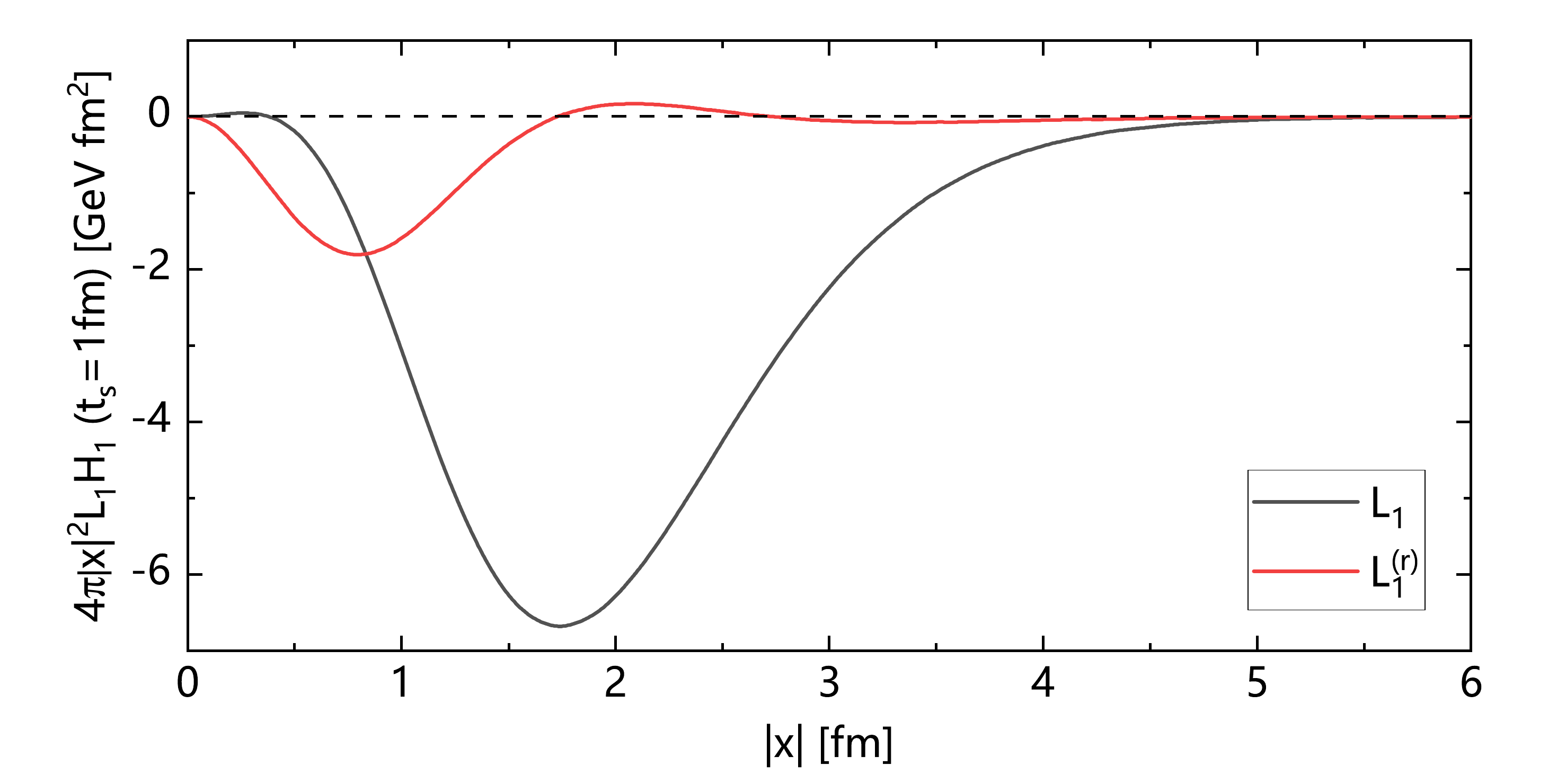}
			\caption{The integrand $4 \pi \abs{\vec{x}}^2 L_1(\vec{x}, t_s) H_1(\vec{x}, t_s)$ as a function of $\abs{\vec{x}}$ estimated by the dipole form factor.}
			\label{fig:L1H1_model}
		\end{center}
	\end{figure}
	
	We set $t_s = 1$ fm for the ground-state saturation, as a result the hadronic function $H_1(\vec{x}, t_s)$ can be estimated by the proton form factor. By using a dipole functional form $G_E(Q^2) = 1/(1+ Q^2 \langle r_p^2 \rangle /12)^2$ with $\sqrt{\langle r_p^2 \rangle} = 0.85$ fm for the form factor, the result of the integrand $4 \pi \abs{\vec{x}}^2 L_1(\vec{x}, t_s) H_1(\vec{x}, t_s)$ as a function of $\abs{\vec{x}}$ is shown in Fig.~\ref{fig:L1H1_model}. We find that the contribution of this integral mainly comes from the range of $1-3$ fm, while the saturation occurs at around $5$ fm, which requires a large spatial volume $L \simeq 10$ fm. We thus set $R_\text{min}$ and $R_\text{max}$ as $1$ fm and $3$ fm. With these parameters, the coefficients can be obtained as $c_0 = -0.17~\mu$eV and $c_r = -93.72~\mu\text{eV}/\text{fm}^2$. As shown in Fig.~\ref{fig:L1H1_model}, we now find the saturation for $L_1^{(r)}(\vec{x}, t_s)$ term occurs at around $2.5$ fm, which means the long-distance contribution is significantly reduced compared to the original $L_1(\vec{x}, t_s)$ term. Finally, the TPE correction to the $2S-2P$ $\mu$H Lamb shift is given by
	\be \label{eq:TPE_split}
	\Delta E_{\text{TPE}} = 0.77~\mu\text{eV} + 93.72~\mu\text{eV}/\text{fm}^2 \langle r_p^2 \rangle - \Delta E^{(\text{lat})}.
	\ee
	Here a minus sign is added due to the TPE correction is only applied to the $nS$-state.
	
\section{Numerical result}
\begin{figure}[htbp]
		\begin{minipage}[t]{0.19\linewidth}
			\centerline{ \includegraphics[width=75pt]{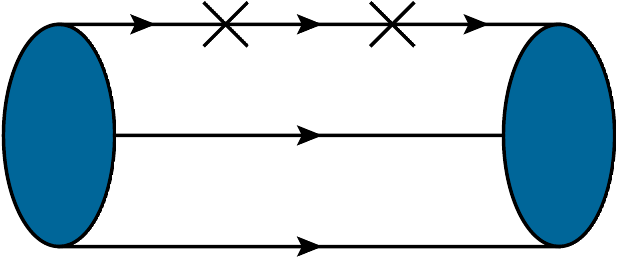} }
			\centerline{Type I}
		\end{minipage}
		\begin{minipage}[t]{0.19\linewidth}
			\centerline{ \includegraphics[width=75pt]{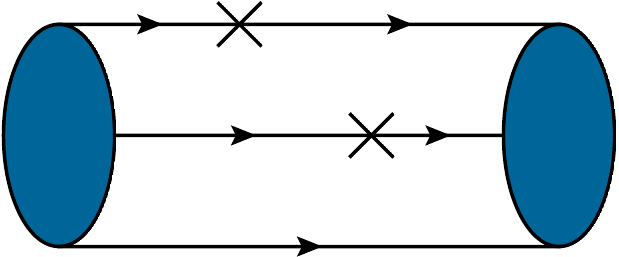} }
			\centerline{Type II}
		\end{minipage}
		\begin{minipage}[t]{0.19\linewidth}
			\centerline{ \includegraphics[width=75pt]{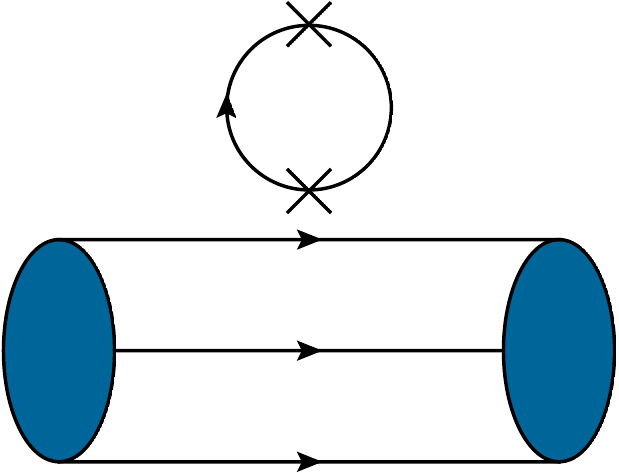} }
			\centerline{Type III}
		\end{minipage}
		\begin{minipage}[t]{0.19\linewidth}
			\centerline{ \includegraphics[width=75pt]{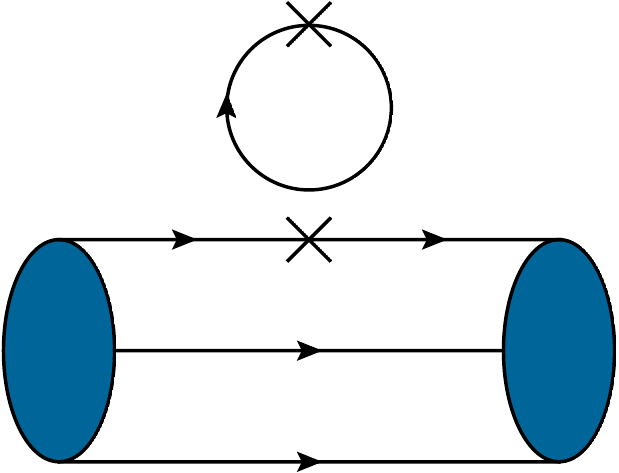} }
			\centerline{Type IV}
		\end{minipage}
		\begin{minipage}[t]{0.19\linewidth}
			\centerline{ \includegraphics[width=75pt]{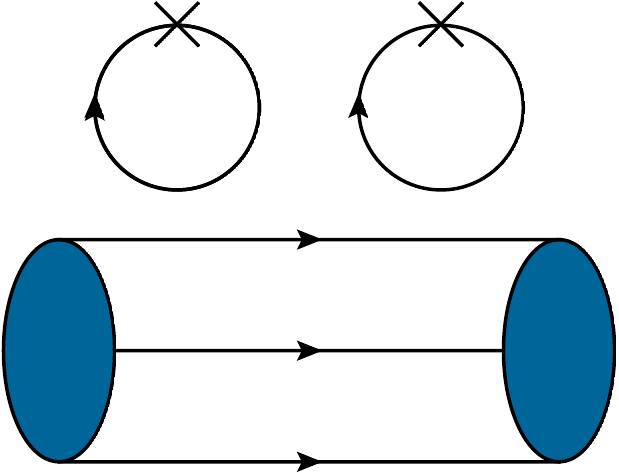} }
			\centerline{Type V}
		\end{minipage}
		\caption{Five types of quark contractions for TPE diagrams. The blob denotes a proton state. Type I: two currents on the same quark line. Type II: two currents on different quark lines. Type III, IV, V: quark disconnected diagrams. The last two types are neglected in this work.}
		\label{fig:contractions}
	\end{figure}

	In this exploratory study, we have used a single gauge ensemble at physical point $m_\pi = 142$ MeV, generated by the RBC and UKQCD Collaborations using $2+1$-flavor domain wall fermion \cite{Blum:2014tka}. The corresponding parameters are listed in Table \ref{tab:ensemble_parameter}. We calculate the four-point correlation function $\sum_{\vec{x}_f, \vec{x}_i} \mathcal{P} \langle \psi_p(\vec{x}_f,t_f) j_\mu(x) j_\nu(y) \psi_p^\dagger(\vec{x}_i,t_i)\rangle$ using the field sparsening technique \cite{Detmold:2019fbk,Li:2020hbj}, with the projection matrix $\mathcal{P} = (1+\gamma_0)/2$ and the time slices chosen as $t_i = \min\{t_x, t_y\} - \Delta t$, $t_f = \max\{t_x, t_y\} + \Delta t$. The $\Delta t$ should be sufficiently large for the proton ground-state saturation, but as the $\Delta t$ increases, the signal-to-noise problem will also be dramatically enhanced. In this exploratory study, $\Delta t$ is chosen to be $2a = 0.39$ fm. There are five types of contractions for the TPE diagrams as shown in Fig.~\ref{fig:contractions}. The first two are quark connected diagrams, while the last three are quark disconnected diagrams. Type IV and Type V are neglected in this work since they vanish in the flavor SU(3) limit. We use the gauge configurations with sufficiently long separation, i.e., each separated by at least 10 trajectories.
	
	\begin{table}[htbp]
		\normalsize
		\centering
		\begin{tabular}{ccccccc}
			\hline
			\hline
			Ensemble & $m_\pi$ [MeV] & $L$ & $T$ & $a$ [fm]&
			$N_{\text{conf}}$ & $\Delta t/a$ \\
			\hline
			24D  & 142 & $24$ & $64$ & $0.1944$ & 131 & 2 \\
			\hline
		\end{tabular}%
		\caption{Ensemble used in this work. We list the pion mass $m_\pi$,  the spatial and temporal extents, $L$ and $T$, the lattice spacing $a$, the number of configurations used $N_{\text{conf}}$, and the time-separation $\Delta t$ used for the ground-state saturation.}
		\label{tab:ensemble_parameter}%
	\end{table}
	
\begin{figure}[htbp]
		\begin{minipage}[t]{0.5\linewidth}
			\centerline{\includegraphics[height=70mm]{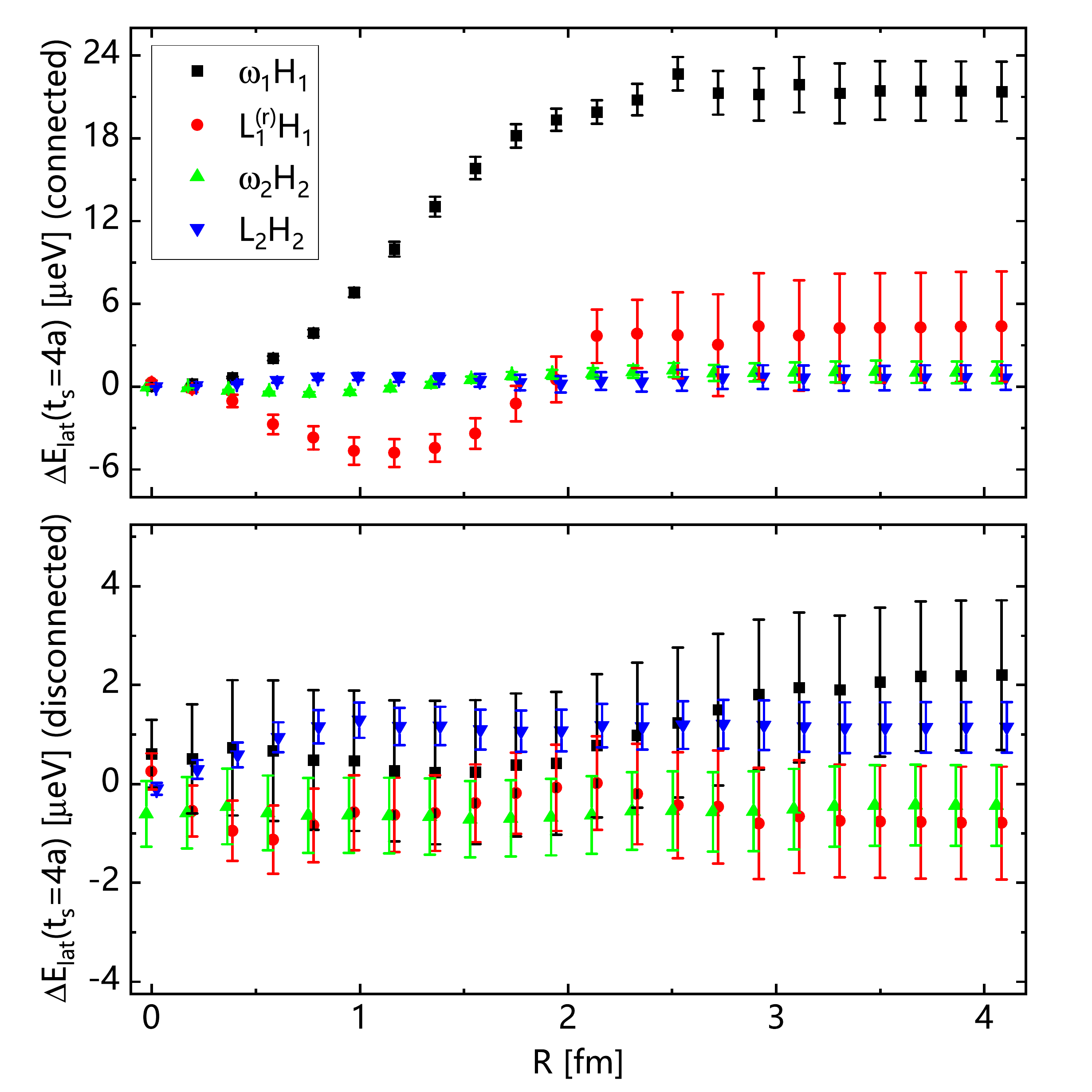}}
		\end{minipage}
		\begin{minipage}[t]{0.5\linewidth}
			\centerline{\includegraphics[height=70mm]{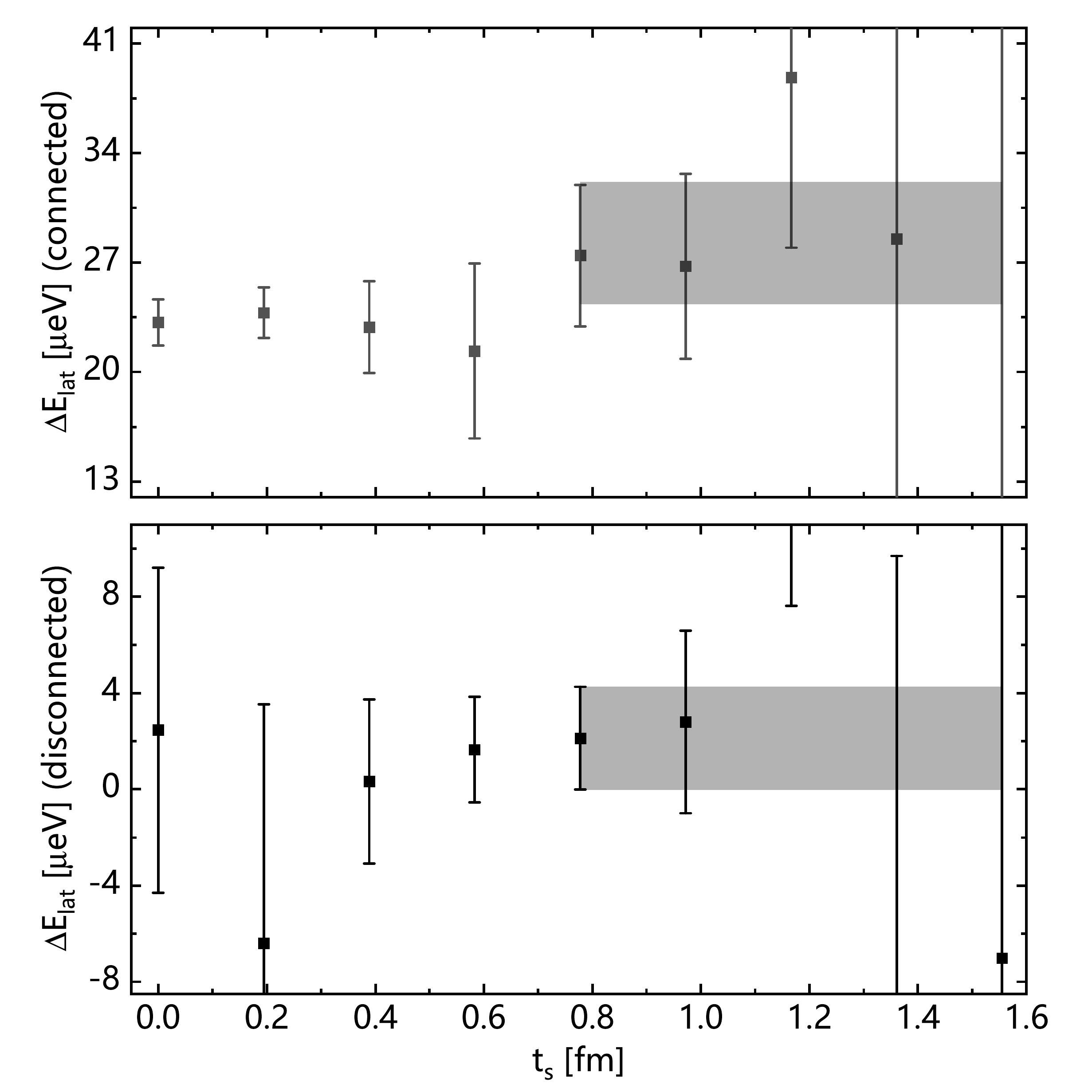}}
		\end{minipage}
		\caption{Left: $\Delta E^{(\text{lat})}$ as a function of the integral range $R$ at $t_s = 4a$. Results from different terms have been slightly shifted for clarity. Right: $\Delta E^{(\text{lat})}$ as a function of $t_s$. For both left and right figure, the upper and lower panels show the results for the connected and disconnected contribution, respectively.}
		\label{fig:TPE_ts}
	\end{figure}

	In practice, the integral in Eq.(\ref{eq:master_formula}) can be performed within a range of $\abs{\vec{x}} < R$ for any choice of time $t_s$. Here we take $t_s = 4a = 0.79$ fm as an example. As shown in the left panel of Fig.~\ref{fig:TPE_ts}, all four terms in the integral is saturated at large $R$ for both connected and disconnected contributions. This indicates the finite-volume effects are well under control in our calculation.
	
	The results of $\Delta E^{(\text{lat})}$ as a function of $t_s$ are shown in the right panel of Fig.~\ref{fig:TPE_ts}. We find a plateau starting from $t_s = 4a = 0.79$ fm. The results are in good agreement with results from data-driven analysis and one magnitude smaller compared to the large $\sim300~\mu$eV discrepancy for the proton radius puzzle. 
	
	The systematic error of our result should mainly come from the excited-states contamination and the lattice discretization error, which requires results from multiple choices of $\Delta t$ and lattice spacing $a$. Further calculations and analyses are in progress.
	
\section{Conclusion}
	We have developed a method to calculate the two-photon exchange correction to the muonic-hydrogen Lamb shift using lattice QCD. We also find that the long-distance contribution can be reduced by adopting an optimized subtraction scheme, hence both the finite-volume effects and nucleon signal-to-noise problem are suppressed. A preliminary result at $m_\pi = 142$ MeV is presented here and it demonstrates the feasibility of our method.
	
	This method can also be extended to other higher-order corrections, including the TPE corrections to other processes such as hyperfine splitting and electron-proton scattering. In a foreseeable future, lattice QCD should be able to give very precise calculations for these quantities, which is crucial for a better understanding and precise measurement of the nucleon internal structure.

\bibliographystyle{JHEP}
\bibliography{ref.bib}

\end{document}